\documentclass[aps,notitlepage, floatfix, prx]{revtex4-1}%

\usepackage{amssymb}
\usepackage{amsfonts}
\usepackage{amsmath}
\usepackage{graphicx}
\usepackage{enumerate}
\usepackage[usenames,dvipsnames]{xcolor}
\usepackage{ulem}%
\providecommand{\U}[1]{\protect\rule{.1in}{.1in}}
\begin{document}
\title{Mechanism for the stabilization of protein clusters above the solubility curve}
\author{James F. Lutsko}
\email{jlutsko@ulb.ac.be}
\homepage{http://www.lutsko.com}
\author{Gr\'{e}goire Nicolis}
\affiliation{Center for Nonlinear Phenomena and Complex Systems, Code Postal 231,
Universit\'{e} Libre de Bruxelles, Blvd. du Triomphe, 1050 Brussels, Belgium}

\begin{abstract}
Pan, Vekilov and Lubchenko[J. Phys. Chem. B 2010, 114, 7620–7630] have proposed that dense stable protein clusters
appearing in weak protein solutions above the solubility curve are composed of
protein oligomers. The hypothesis is that a weak solution of oligomer species
is unstable with respect to condensation causing the formation of
dense, oligomer-rich droplets which are stabilized against growth by the
monomer-oligomer reaction. Here, we show that such a combination of
processes can be understood using a simple capillary model yielding analytic expressions for the cluster properties which can be used to interpret experimental data. We also construct
a microscopic Dynamic Density Functional Theory model and show that it is consistent
with the predictions of the capillary model. The viability of the mechanism is
thus confirmed and it is shown how the radius of the stable clusters is
related to physically interesting quantities such as the monomer-oligomer rate constants.
\end{abstract}

\date{\today }
\maketitle

\section{Introduction}
Proteins in solution demonstrate a surprisingly rich variety of phenomena,
many of which have biological implications. These include liquid-liquid
separation (e.g. protein rich droplets forming in a weak solution),
crystallization and gelation\cite{Gunton2007}. Perhaps most surprising is the relatively
recent discovery of the existence of protein clusters having a typical, stable
size and long life time that have been found to exist under a wide variety of
conditions including those for which no condensed phases are stable\cite{VekilovClusters_Evidence1,VekilovClusters_Evidence2,VekilovClusters_Evidence3}. There is abundant evidence that these clusters play an important role
in protein crystallization\cite{VekilovCGDReview2004,LN,MikePNAS} and it has been suggested that they play a
similar role in the formation of protein aggregates that are actors in
pathologies such as hemoglobin polymers in sickle cell anemia and fibrils of
misfolded proteins that underlie various neurological disorders\cite{VekilovClusters1}. Thus,
understanding their origin and nature is of  importance for both
fundamental and practical reasons. 

The existence of stable clusters anywhere in the phase diagram is mysterious. A subcritical droplet of a condensed phase should, by definition, evaporate whereas a supercritical droplet should grow until all available material is incorporated. Multiple droplets can compete with one another for the available material slowing down the growth process during the so-called ripening stage, until only very large droplets remain. Coelescence of droplets can also contribute to this outcome. However, the experimental evidence indicates that for the protein clusters, any such ripening ends after a finite time leaving a stable population of droplets much smaller than expected according to classical scenarios\cite{VekilovClusters3}. 

Recently, Vekilov and co-workers proposed that the stable protein
clusters might not be composed of the native protein but, rather, of some
complex formed from them such as an oligomer or a mis-folded monomer\cite{VekilovClusters1,VekilovClusters2,VekilovClusters3}. Their idea was that in the original, weak protein solution the new species  is in equilibrium with the
protein monomer but that the phase diagram of the new species is such that a
condensed phase is favored so that  super-critical clusters (e.g.  oligomer-rich  droplets) can form. However, since the density of of the secondary species
within the clusters  would be well above the concentration for chemical
equilibrium with the monomer, there would be a tendency for the secondary
species  to convert back to monomers within the cluster thus impeding its
growth. If one adds the assumption that excluded volume effects prevent a high
monomer concentration within the droplet (which could also lead to a stable
chemical equilibrium between the two species) then a possible mechanism for
stabilization is apparent. Although this idea has motivated further
experimental work, little has been done to formalize it theoretically. The
goal of this paper is to do so at two levels. First, the stabilization problem
will be considered phenomenologically using concepts from Classical Nucleation
Theory(CNT)\cite{Kashchiev}. This will result in simple analytic expressions for the size
of the stable clusters as functions of the properties of the original solution
and of the concentration and pressure of the secondary species. This analytic relation opens the door to the
determination of these properties from experiment.
Our second contribution is the formulation of  a detailed Dynamic Density Functional Theory (DDFT) model  based on the same assumptions and used to confirm the
phenomenological predictions while providing a more fundamental means of
investigating the nature of the clusters. 

\section{Phenomenology}
For the sake of concreteness, we will assume that the secondary population is
composed of dimers. All of the subsequent development can be trivially adapted
to other possibilities. We then begin by postulating a simple mass-balance
reaction model for the conversion of monomers into dimers and vice-versa.
Calling the monomer number density $n_{1}$ and the dimer density $n_{2}$ this
takes the form%
\begin{align}
\frac{dn_{1}}{dt} &  =-2k_{1}n_{1}^{2}+2k_{2}n_{2}\\
\frac{dn_{2}}{dt} &  =k_{1}n_{1}^{2}-k_{2}n_{2}\nonumber
\end{align}
where the factors of two ensure that the total number of protein molecules,
$n=n_{1}+2n_{2}$ within any small volume element, is conserved in the absence of spatial inhomogeneities. The square of the monomer density occurs
because two monomers must meet to form a dimer. This gives a relation between
the equilibrium densities of $k_{1}n_{1}^{(eq)2}=k_{2}n_{2}^{\left(
eq\right)  }$. The rate equations can be solved exactly and it is found that
the non-conserved difference $n_{1}-2n_{2}$ relaxes exponentially at long times with time
constant $\sqrt{k_{2}^{2}+8nk_{1}k_{2}}$.

Now, let us consider a pure solution of dimers and we assume that conditions
are such that the fluid nucleates a dense phase. In the capillary
approximation used in CNT, it is assumed that the density inside a cluster
having radius $R$ is constant, $n_{2}\left(  r<R\right)  =n_{2}^{\left(
0\right)  }$ and equal to the density of the homogeneous, condensed phase, while the density outside the cluster, $n_{2}\left(  r>R\right)
=n_{2}^{\left(  \infty\right)  }$ is also constant. In this case, the rate of growth of a
sufficiently large supercritical cluster is, under the diffusion-limited
conditions expected to dominate for macromolecules in solution, given by \cite{Lutsko_JCP_2012_1}
\begin{equation}
\frac{dR}{dt}=aR^{-1},a=Dn_{2}^{\left(  \infty\right)  }\frac{\beta P\left(
n_{2}^{\left(  0\right)  }\right)  -\beta P\left(  n_{2}^{\left(
\infty\right)  }\right)  }{\left(  n_{2}^{\left(  0\right)  }-n_{2}^{\left(
\infty\right)  }\right)  ^{2}}\label{R}%
\end{equation}
where $D$ is the tracer diffusion constant for a dimer molecule in solution,  $P\left(
n_{2}\right)  $ is the pressure for the dimers at density $n_{2}$ and
$\beta=1/k_{B}T$ with $T$ the temperature and $k_{B}$ Boltzmann's constant..
This gives the classic result $R \sim t^{1/2}$. 

Now, let us consider the effect of adding monomers to the picture. Outside the
cluster, the monomers and dimers will reach equilibrium so we have that
$k_{1}n_{1}^{(\infty)2}=k_{2}n_{2}^{\left(  \infty\right)  }$. We assume that
the monomers and dimers have no interaction aside from excluded volume
effects. In this case adding monomers to the cluster raises its free
energy  so that one expects the monomers to be expelled by diffusion leading
to the hypothesis that the monomer concentration inside the cluster is very
low,  $n_{1}^{\left(  0\right)  }\simeq0$. Clearly, the realization of this
condition will depend on diffusion being sufficiently fast compared to the
rate of production of the monomers. In terms of the dimer concentration within
the cluster, the net effect (conversion of dimers to monomers and expulsion of
the monomers) is a simple extinction reaction that lowers the total number of
dimers, $N_{2} = \frac{4 \pi}{3}n_{2}R^{3}$,  according to $dN_{2}/dt=-k_{2}N_{2}$. Since the free energy of the
cluster will be minimized by maintaining a dimer density near that of the
thermodynamically stable condensed phase, this leads to a reduction in the
size of the cluster given by $dR/dt=-k_{2}R/3$.  

The combined effect of the reaction and of diffusion gives an evolution
equation for the radius of the form
\begin{equation} \label{Req}
\frac{dR}{dt}=aR^{-1}-k_{2}R/3.
\end{equation}
In this simple relation, the term driving growth scales more weakly than the
term opposing growth which is the opposite of what happens in classical
nucleation theory. As a consequence, the dynamics are reversed: small clusters
tend to grow while large clusters tend to shrink until the cluster reaches a stable, stationary size as is reflected in the exact solution to Eq.(\ref{Req}),
\begin{equation} \label{Rcap}
R^{2}\left(  t\right)  =R_{s}^{2}+\left(  R^{2}\left(  0\right)  -R_{s}%
^{2}\right)  e^{-\frac{2k_{2}}{3}t}, \,\,\, R_{s}=\sqrt{\frac{3a}{k_{2}}}.%
\end{equation}

These expressions link accessible experimental quantities such as the cluster size and the rate of relaxation of the system to the parameters governing the model.
In particular, they in principle
give experimental access to the rate constants since one expects
the exterior dimer concentration, $n_{2}^{\left(  \infty\right)  }$, to be in
equilibrium with the monomer concentration outside the droplet ($k_{1}%
n_{1}^{(\infty)2}=k_{2}n_{2}^{\left(  \infty\right)  }$) so that measurement
of the respective concentrations, together with knowledge of $k_{2}$, allows
the determination of $k_{1}$ and thus complete characterization of the
reaction between the two species.

\section{Microscopic model}
To test these ideas, we now describe a microscopic model that incorporates the
growth of a super-critical droplet and the excluded volume interaction of the
monomer and dimer species. Our approach is based on Dynamic Density Functional
Theory (DDFT) which is commonly used to describe the dynamics of over-damped
systems (such as colloids and macromolecules in solution) under conditions
such that thermal fluctuations may be ignored\cite{MarconiTarazona,EvansArcher,lutsko:acp}. In DDFT, the fundamental
quantity is the time-dependent local density (or equivalently, concentration)
$n\left(  \mathbf{r};t\right)  $. The diffusion-limited growth of a
super-critical droplet in a pure solution of dimers (i.e. with no monomers
present) is governed by%
\begin{equation}
\frac{dn_{2}\left(  \mathbf{r};t\right)  }{dt}=D_{2}\mathbf{\nabla\cdot}%
n_{2}\left(  \mathbf{r};t\right)  \mathbf{\nabla}\frac{\delta F\left[
n_{2}\right]  }{\delta n_{2}\left(  \mathbf{r};t\right)  }%
\end{equation}
where $D_{2}$ is the tracer diffusion constant for the dimers. The free energy
functional will be taken to have the squared-gradient form\cite{Evans1979,lutsko:acp}%
\begin{equation}
F\left[  n_{2}\right]  =\int\left\{  f_{2}\left(  n_{2}\left(  \mathbf{r}%
;t\right)  \right)  +\frac{1}{2}g_{2}\left(  \nabla n_{2}\left(
\mathbf{r};t\right)  \right)  ^{2}\right\}  d\mathbf{r}%
\end{equation}
where $f_{2}\left(  n_{2}\right)  $ is the Helmholtz free energy per unit
volume for a homogeneous fluid at density $n_{2}$ and $g_{2}$ is a constant
that can be calculated from the interaction potential\cite{Lutsko2011a}. In the
following, the dimers will be described generically using a Lennard-Jones
interaction potential in which case good parameterizations are available in
the literature\cite{JZG}. In the limit of low densities, the gradient term
is negligible and the Helmholtz free energy goes to the ideal gas form
$f_{2}\left(  n_{2}\right)  \rightarrow f^{\left(  id\right)  }\left(
n_{2}\right)  =n_{2}\ln n_{2}\Lambda^{3}-n_{2}$ so that the left hand side of
the DDFT equation reduces to $D_{2}\nabla^{2}n_{2}$ , i.e. it becomes the
diffusion equation. Thus, one may think of DDFT as a generalization of the
diffusion equation that accounts for particle interactions.

To generalize to two species, the free energy functional is replaced by one
depending the local densities of both species, $F\left[  n_{1},n_{2}\right]
$, and a second DDFT equation is included for $n_{1}$. In the present case, we
must also include the chemical reactions thus giving%
\begin{align}
\frac{dn_{1}\left(  \mathbf{r};t\right)  }{dt} &  =  D_{1}\mathbf{\nabla\cdot
}n_{1}\left(  \mathbf{r};t\right)  \mathbf{\nabla}\frac{\delta F\left[
n_{1},n_{2}\right]  }{\delta n_{1}\left(  \mathbf{r};t\right)  }   -2k_{1}%
n_{1}\left(  \mathbf{r};t\right)  ^{2}+2k_{2}n_{2}\left(  \mathbf{r};t\right)
\\
\frac{dn_{2}\left(  \mathbf{r};t\right)  }{dt} &  =  D_{2}\mathbf{\nabla\cdot
}n_{2}\left(  \mathbf{r};t\right)  \mathbf{\nabla}\frac{\delta F\left[
n_{1},n_{2}\right]  }{\delta n_{2}\left(  \mathbf{r};t\right)  }  +k_{1} n_{1}\left(  \mathbf{r};t\right)  ^{2}-k_{2}n_{2}\left(  \mathbf{r};t\right)
\nonumber
\end{align}
In principle, for a non-ideal system we should replace the concentrations occurring in the chemical reaction terms by the corresponding activities. Here, we keep the simple form given above for the sake of comparison to the phenomenological model and defer further discussion of this point to the Conclusions.

Finally, the form of the free energy functional must be specified. Since the
monomers are supposed to be above their critical point, we simply
treat them as hard spheres with hard-sphere diameter $d$ so as to account for
excluded volume effects. The final form we employ is
\begin{align} \label{Model}
F\left[  n_{1},n_{2}\right]   &  =\int\left\{  f_{hs}\left(  n_{1}\left(
\mathbf{r};t\right)  ;d\right)  +\frac{1}{2}g_{1}\left(  \nabla n_{1}\left(
\mathbf{r};t\right)  \right)  ^{2}\right\}  d\mathbf{r}\\
&  +\int\left\{  f_{LJ}\left(  n_{2}\left(  \mathbf{r};t\right)  \right)
+\frac{1}{2}g_{2}\left(  \nabla n_{2}\left(  \mathbf{r};t\right)  \right)
^{2}\right\}  d\mathbf{r}\nonumber\\
&  +\int\left\{  f_{hs}^{(ex)}\left(  n_{1}\left(  \mathbf{r};t\right)
+n_{2}\left(  \mathbf{r};t\right)  ;d\right)  -f_{hs}^{(ex)}\left(
n_{1}\left(  \mathbf{r};t\right)  ;d\right)  -f_{hs}^{(ex)}\left(
n_{2}\left(  \mathbf{r};t\right)  ;d\right)  \right\}  d\mathbf{r}\nonumber
\end{align}
The third line accounts for the mutual excluded volume interaction of the two
species: we treat both as hard spheres of diameter $d$ and replace their
individual hard-sphere contributions to the excess free energy by one
dependent on the sum of the local densities.\ (Note that the excess part of
the free energy is just $f^{\left(  ex\right)  }=f-f^{\left(  id\right)  }$:
we only replace the excess part because the ideal contributions are already
accounted for.) If either density is zero, this interaction term vanishes. Of
course, a dimer with twice the mass of a monomer and the same density would
have a diameter about $25\%$ larger but for simplicity we ignore this
relatively small difference. Similarly, we take $g_{1}=g_{2}=g_{LJ}$ and
$D_{1}=D_{2}$ since we expect the differences in these coefficients to be of
no physical importance. A final simplification is that we do not include a
cross term involving the gradients. This model is a generalization of the model used by Huberman to discuss the appearance of striations in a reacting system\cite{Huberman}. Huberman's model was constructed in the approximation of a single active reactant with an autocatalytic chemical reaction out of equilibrium. Here, the presence of two species participating in an equilibrium reaction is fully accounted for. This necessarily requires adding an additional contribution to the free energy and, most importantly, the third line in Eq.(\ref{Model}) which accounts for the most basic excluded-volume interaction of the two species. Note that in this setting the conservation condition $n_{1}+2n_{2} = \mathrm{const.}$ no longer holds locally.

The Lennard-Jones potential introduces a length scale, $\sigma$, and an energy
scale $\varepsilon$. In the following, temperature will be reported in the
scaled units $T^{\ast}=k_{B}T/\varepsilon$ and all lengths will be scaled by
$\sigma$. We also take the hard-sphere diameter $d=\sigma$:\ typical
prescriptions such as Barker-Henderson\cite{BarkerHend} change this by a few percent but
for present purposes this difference is unimportant. A time scale, $\tau$, is
introduced such that $D_{2}\tau/\sigma^{2}=1$. After scaling, the available
parameters are the monomer background density, the dimer supersaturation, the
scaled temperature and the scaled reaction coefficient $k_{1}^{\ast}$. The
dimer reaction constant is determined via the equilibrium condition
$k_{1}^{\ast}n_{1}^{\left(  \infty\right)  \ast2}=k_{2}^{\ast}n_{2}^{\left(
\infty\right)  \ast}$. We report results here for $T^{\ast}=0.8$ and
supersaturation $n_{2}^{\left(  \infty\right)  }/n_{2}^{\left(  coex\right)
}=2$ where $n_{2}^{\left(  coex\right)  }$ is the vapor density at coexistence
at this temperature. Under these conditions the density in the vapor is
$n_{2}^{\left(  \infty\right)  \ast}=0.012$ and in the condensed phase
$n_{2}^{\ast}=0.85$. The background monomer density is taken to be $5$ times
that of the dimer phase. In reality, this ratio is thought to be much greater\cite{VekilovClusters1}
 but the computational cost of the calculation increases with this ratio so
our choice represents a compromise. The only remaining parameter is
$k_{1}^{\ast}$ which is discussed below.

Our calculations were performed assuming spherical symmetry with boundary conditions appropriate for an open system (see Supplementary Material for technical details). We began by
locating the critical cluster for the pure dimer system. With the chosen
parameters, this has radius $R_{c}^{\ast}=5.2$. We then make this supercritical by
increasing its radius an amount $\Delta R$ and then adding in the monomers. Further details are given in the Supplementary Material as are details of the numerical algorithms used to integrate the DDFT equations. Also discussed there are the question of the definition of the radius to use for comparing the capillary model to the DDFT and the agreement between the two theories for the case of the growth of a super-critical droplet in a single-component system. 

\begin{figure}
[ptb]\includegraphics[angle=0,scale=0.35]{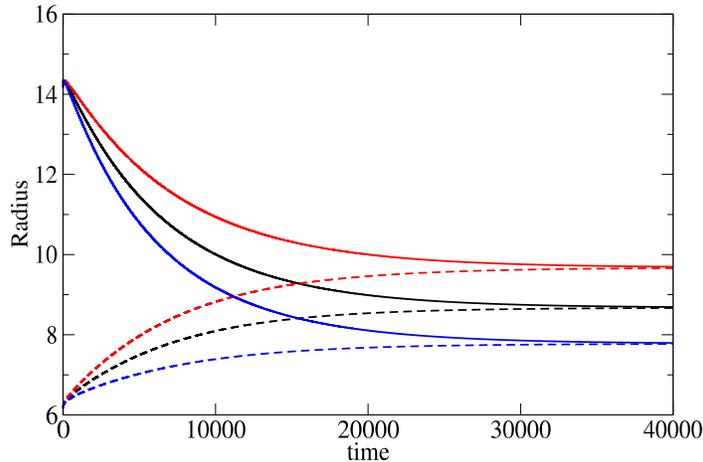}
\caption{Behavior of the cluster radius as a function of time (both in dimensionless units) for three different values of the reaction parameter, $k_{1}^{*} = 8.75 \times 10^{-4}$ (upper curve), $7.5 \times 10^{-4}$ (middle curve) and $10^{-3}$ (lower curve). In each case, two initial configurations are used: one with a small initial displacement of the critical cluster, and one with a large initial displacement. In all three cases, both initial conditions lead to the same final cluster radius thus demonstrating the stability of the final cluster.}
\label{fig_convergence}
\end{figure}

\begin{figure}
[ptb]\includegraphics[angle=0,scale=0.35]{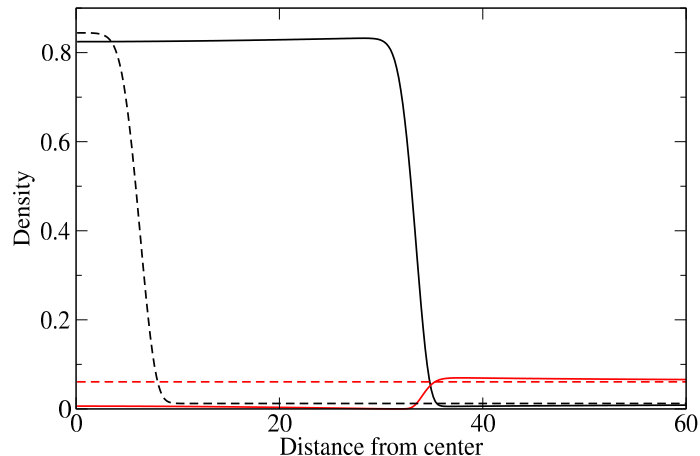}
\caption{Structure of the stable cluster for $k_{1}^{*} = 7.5 \times 10^{-5}$. The density (concentration) of the monomer species (solid red line) and the dimer species (solid black line) is shown as functions of distance from the center of the cluster. The initial condition is also shown using dashed-lines.}
\label{fig_structure}
\end{figure}

The evolution of the cluster radius for three different values of the reaction
constant is shown in Fig. \ref{fig_convergence}. In each case, two initial displacements are used:
an "under" displacement of one unit (broken lines) and an "over" displacement
of 9 units (full lines). The fact that the under- and over-displaced clusters
evolve to the same final cluster is strong empirical evidence for the
stability of the final cluster. The structure of the stable cluster is shown
in Fig. \ref{fig_structure}  where it can be seen that most of the monomer species is expelled
from the cluster except in the interfacial region.

\begin{figure}
[ptb]\includegraphics[angle=0,scale=0.35]{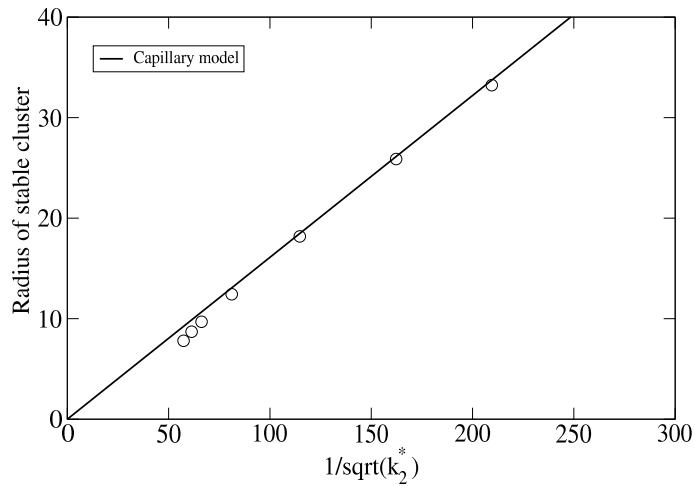}
\caption{Predicted stable radius from the capillary model, Eq.(\ref{Rcap}), compared to the results of numerical integration of the DFT model (symbols) as a function of $1/\sqrt{k_{2}^{*}}$. With these variables, the prediction is simply a straight line.}
\label{fig_k}
\end{figure}

The scaling relation between the stable radius and the reaction constant
$k_{2}^{*}$ predicted by the capillary model, Eq.(\ref{Rcap}), is tested
against the numerical DFT results in Fig. \ref{fig_k}. For lower values of the
dimensionless reaction coordinate, there are significant deviations as is to
be expected since the capillary model is only accurate for large clusters. As
the reaction rate decreases, and the size of the stable cluster increases,
convergence to the prediction is evident.

\section{Conclusions}
We have shown that super-critical clusters, which would otherwise be unstable
with respect to growth, can be stabilized by means of the combined effect of
diffusion and a chemical reaction. Diffusion - driven by thermodynamics -
leads to the purification of the cluster so as to lower its free energy. The
purified cluster is then in turn subject to degradation due to the conversion
of the dimer species to monomers. This dynamic process can be successfully
described by a simple capillary model as well as more systematically
investigated by means of a microscopic Dynamic Density Functional Theory
model. The two approaches were shown to be in agreement. While these results were necessarily achieved for specific choices of molecular interactions and, particularly, for a specific chemical reaction, it is clear that the arguments may be trivially adapted to other choices. 

The microscopic DDFT model presented here is a natural generalization of the standard reaction-diffusion model used to describe chemical reactions in spatially extended systems. 
The crucial element in our formulation of this generalization is the free energy functional and particularly the interaction term given in the third line of Eq.(\ref{Model}). The free energy contribution can be viewed as giving rise to a density-dependent diffusion constant which, for the condensed phase, is negative thus driving growth of a cluster rather than its diffusive evaporation as in ordinary diffusion. The interaction term in the free energy is critical in that it leads to a monomer diffusion constant that increases with increasing density of dimers thus causing expulsion of the monomers from the dimer cluster. This leads to a locally frozen nonequilibrium steady state in which a current of dimers flows into the cluster where they are converted to monomers and expelled in the form of a corresponding outward current. In this state growth of the droplet and the conversion of species are mutually quenched. Since such a nonequilibrium state cannot persist indefinitely without a driving force (due e.g., to mode-coupling effects not considered in the over-damped limit used here\cite{Lutsko_JCP_2012_1}), the clusters are not expected to be stable indefinitely. Furthermore, shape fluctuations are also likely to prove destabilizing since any deviation from a spherical shape will lower the thermodynamic driving force for growth and potentially lead to irreversible shrinking of the cluster to a size below the critical radius. Finally, as mentioned already earlier, in  the results presented here the system is actually treated as an \emph{open} system spatially infinite, continually replenished by monomers  and  coupled implicitly to an infinite solvent\cite{Lutsko_JCP_2012_1}  that acts as a reservoir. This further postpones the establishment of a global equilibrium throughout.

We conclude with several observations concerning this mechanism. First, there is no constraint on the free energy of the stable droplet since
the only requirement is that it be larger than the critical cluster. It could
therefore have a free energy nearly as high as that of the critical cluster
(leading to a relatively low number of such droplets in equilibrium) or it
could have an arbitrarily low free energy (leading to a large population).

Second, we have assumed that when the reaction removes dimers the
density of the cluster remains constant so that the net effect is that the
cluster shrinks in size. This only makes physical sense if the reaction is in
some sense slow compared to the process of removing monomers from the cluster
(i.e. diffusion). Were this not the case, monomers would quickly build up
within the cluster and poison it leading to its collapse.

Third, we note the generality of the mechanism leading to a stable cluster
with a characteristic size: a force driving growth that scales more slowly
than a force opposing growth. Regardless of the mechanisms giving rise to the
forces, these are the required elements. Clusters in other systems could be
stabilized by some other combination of growth-promoting and -opposing forces
provided the relative scaling satisfies this rule.

Fourth, one can contrast this mechanism to that stabilizing vesicles. The
latter consist of a volume with amphiphilic molecules arranged on its surface
so that their hydrophobic parts are inside the volume, shielded from water,
while their hydrophilic parts are on the outside of the surface, exposed to
the water. Within the vesicle could be void, more of the apmphiphilic
molecules or some other substance. If the interior has a higher free energy
than the solution, the vesicle can be stabilized in the same manner as
proposed above: the surface dominates the free energy of small vesicles
leading to growth while the volume dominates large vesicles leading to
dissolution. However, in the case of vesicles there is another factor: such a
system can increase its surface to volume ratio, and hence decrease its free
energy, by becoming non-spherical ( by becoming flat, in the extreme limit).
In our case, the free energy is minimized by a spherical shape so that the
mechanism favors the formation of spherical clusters.

Fifth, there is no scope within this model for ripening: i.e. the growth of larger clusters at the expense of smaller ones. Something like ripening has in fact been reported by Li et al.\cite{VekilovClusters3} albeit with the unusual feature that the ripening stops while there is still a finite population of clusters. If the present model were correct, this ``ripening'' would have to be reinterpreted: perhaps as a slowly relaxing transient. As stated above, the reaction must be slow compared to diffusion, as is reflected in the small dimensionless reaction constants used in our work, and this could simply result in very slow dynamics for the entire system. Alternatively, it is possible that the dimer to monomer reaction is suppressed within the cluster (due to the high free energy barrier involved in removing a dimer from the condensed phase) and that the reaction is most productive only in the boundary of the cluster (where the dimer is in an energetically unfavorable state). In this case, the reaction term in Eq.(\ref{Req}) would be a constant rather than scaling as $R$ (in fact, $R$ would be replaced by the characteristic width of the boundary region) and this would lead to algebraic rather than exponential relaxation of the cluster to its stable size. To capture this, the model could be modified by replacing the concentrations in the rate equations with more general expressions involving the chemical affinities. Such an algebraic dynamics combined with small reaction constants could well give transients that decay very slowly and could therefore be interpreted as a transient ripening. 

This is related to our sixth and final point. We mentioned above that for consistency, we should replace the concentrations appearing in the chemical reaction kinetics by the corresponding activities, $n_{i}(r,t) \rightarrow n_{i}^{(0)}\exp(\beta \mu_{i}(r,t)-\beta \mu_{i})$ where the local chemical potential is $\mu_{i}(r,t) = \frac{\delta F}{\delta n_{i}(r,t)}$ and where $\mu_{i}$ is the chemical potential for species $i$ in the homogeneous system in which $n_{i}(r,t) = n_{i}^{(0)}$. This has not been used in the present work  in order to explore the consistency of the simple capillary model with the microscopic model in the case that the relation between the two is most straightforward. We speculate that the effect of the use of the activities will be a suppression of the dimer to monomer reaction within the cluster and an enhancement of the importance of the reaction in the interfacial region, therefore possibly leading to the scenario alluded to above of an algebraic rather than exponential relaxation. Preliminary calculations using the activities supports this and the issue will be discussed more fully in a future publication. 


\begin{acknowledgments}
This work was supported in part by the European Space Agency under contract
number ESA AO-2004-070. The authors thank Peter Vekilov for numerous conversations on this subject and Pierre Gaspard for useful comments concerning the model. 
\end{acknowledgments}

\bibliography{protein_clusters}
\end{document}